\documentclass[12pt]{article}
\usepackage[english]{babel}
\usepackage{amsmath,amsthm,amssymb,amsfonts}
\usepackage{color}

\usepackage{cite}
\newtheorem{thm}{Theorem}[section]
\newtheorem{remark}[thm]{Remark}
\newtheorem{prop}[thm]{Proposition}

\begin{document}

\title{Hierarchy of coupled Burgers-like equations induced by conditional symmetries}
\author{M.~Gorgone, F.~Oliveri and E.~Sgroi\\
\ \\
{\footnotesize Department of Mathematical and Computer Sciences,}\\
{\footnotesize Physical Sciences and Earth Sciences, University of Messina}\\
{\footnotesize Viale F. Stagno d'Alcontres 31, 98166 Messina, Italy}\\
{\footnotesize matteo.gorgone@unime.it; francesco.oliveri@unime.it; emanuele.sgroi@unime.it}
}

\date{Published in \emph{Zeitschrift f\"ur Angewandte Mathematik und Physik (ZAMP) \textbf{76}, 8 (2025).}}

\maketitle

\begin{abstract} 
It is known that $Q$-conditional symmetries of the classical Burgers' equation express in terms of three functions satisfying a coupled system of Burgers-like equations. The search of conditional symmetries of  this system leads to a system of five coupled Burgers-like equations. Using the latter system as a starting point, and iterating the procedure, an infinite hierarchy of systems made of an odd number of coupled Burgers-like equations can be conjectured. Moreover, starting from a pair of Burgers-like equations, a similar hierarchy of systems made of an even number of coupled Burgers-like equations may arise. We prove that these two infinite hierarchies can be unified, and each element of the hierarchy arises from the nonclassical symmetries of the previous one. Writing a generic element of this hierarchy as a matrix Burgers' equation, the existence of the matrix Hopf-Cole transformation allows for its linearization and the determination of its solutions. Finally, it is shown that each element of the hierarchy possesses a five-dimensional Lie algebra of classical point symmetries. Though these Lie algebras are realized in manifolds with different dimensionality, they are all isomorphic.
\end{abstract}

\noindent
\textbf{Keywords.} Lie symmetries; Conditional symmetries; Hierarchy of coupled Burgers-like equations;
Matrix Burgers's equation; Matrix Hopf-Cole transform.

\noindent
\textbf{Mathematics Subject Classification (2010).} {35K10 - 58J70 - 58J72}

\section{Introduction}
Lie groups of continuous transformations,  originally introduced in the nineteenth century by Sophus Lie \cite{Lie1,Lie2}, provide a unified and powerful framework for the investigation of  differential equations. Continuous transformations are characterized by their infinitesimal generators, and establish a diffeomorphism on the space of independent and dependent variables, taking solutions of the equations into other solutions (see, for instance, \cite{Ovsiannikov,Olver1,Bluman-Kumei,Ibragimov:CRC}); any transformation of the independent and dependent variables naturally induces a transformation of the derivatives. The Lie group of point transformations leaving a differential equation invariant is obtained by solving a linear system of determining equations for the components of the infinitesimal generators.
This process is completely algorithmic; nevertheless, it usually involves a lot of cumbersome and tedious calculations. The current availability  of many powerful Computer Algebra Systems (CAS) (either commercial or open source) greatly helped the application of Lie group methods, so that most of the needed algebraic manipulations can now be done quickly and often automatically. 
In fact, many specific packages for performing symmetry analysis of differential equations are currently available in the literature~\cite{Hereman1997,Baumann2000book,CarminatiWu2000,%
ButcherCarminati2003,Cheviakov2007,Steeb2007book,Cheviakov2010,%
Sade2010,JeffersonCarminati2013}.
Anyway, the calculations reported in this paper are made by using \texttt{ReLie}  \cite{Oliveri:relie}, a package written in the open source CAS Reduce \cite{Reduce}. 

Lie's theory allows for  the development of systematic procedures for integrating by quadrature (or, at least, lowering the order) of ordinary differential equations \cite{BlumanAnco}, or determining invariant solutions of initial and boundary value problems \cite{Olver-Rosenau-2}, or deriving conserved quantities \cite{Noether,BlumanCheviakovAnco}, or building relations between different differential equations that turn out to be equivalent \cite{BlumanCheviakovAnco,Oliveri:Symmetry,Oliveri:quasilinear,GorgoneOliveri2}.

Along the years, many extensions and generalizations of Lie's approach have been introduced, and a wide set of applied problems has been deeply analyzed and solved. Here, we limit ourselves to consider conditional symmetries whose origin can be thought of as originated in 1969 with the paper by Bluman and Cole 
\cite{Bluman-Cole-1969} on the \emph{nonclassical method}, where the continuous symmetries mapping the manifold characterized by the linear heat equation and the invariant surface condition (together with its differential consequences) into itself have been considered. Later on, Fuschich \cite{Fuschich,Fuschich-Tsyfra} introduced the general concept of conditional invariance. The basic idea consists in replacing the conditions for the invariance of a differential equation by the weak conditions of the invariance of the combined system made by the original equation, and some differential constraints. When the differential constraints are given by the invariance surface conditions and their differential consequences, we have what are usually referred to as $Q$-conditional symmetries. Using this approach, one has  fewer determining equations, with the additional complication that they are nonlinear \cite{Levi-Winternitz,Clarkson-Kruskal,PucciSaccomandi,Arrigo,Olver1994,Olver-Vorobev,Saccomandi}. 
In many cases, the results allow for the introduction of non-trivial  reductions of partial differential equations leading to the determination of wide classes of exact solutions in many applied problems \cite{Cherniha-JMAA2007,Cherniha-JPA2010,Cherniha-Davydovych}. 
Some authors \cite{Kunzinger-Popovich} do not recognize the status of symmetry to nonclassical symmetries, and prefer to use the term \emph{reduction operators}. 
Nevertheless, our primary aim in this paper is not concerned with the use of conditional symmetries to characterize reductions of partial differential equations and determine exact solutions. In fact, we are interested in deriving an infinite hierarchy of systems of second order partial differential equations with a common underlying structure. 

In this paper, within the framework of $Q$-conditional symmetries, we start with the classical Burgers' equation, say 
\begin{equation}
\label{classicalburgers}
u_{,t}+uu_{,x}-u_{,xx}=0
\end{equation}
(the subscripts ${}_{,t}$ and ${}_{,x}$ denoting partial derivatives with respect to $t$ and $x$, respectively), 
or with a system of two coupled Burgers-like equations, say 
\begin{equation}
\label{coupleburgers}
\begin{aligned}
&u_{,t}+uu_{,x}-u_{,xx}+v_{,x}=0,\\
&v_{,t}+vu_{,x}-v_{,xx}=0,
\end{aligned}
\end{equation}
and prove that an infinite hierarchy of coupled Burgers-like equations can be suitably generated. 
The Burgers' equation represents the simplest model equation
suitable to describe wave propagation phenomena when there is a balance
between linear evolution, quadratic nonlinearity and viscous diffusion. Apart
from its own interest since it models various physical situations \cite{Crighton}, the
Burgers' equation is remarkable because of its linearization through the well
known Hopf-Cole transformation \cite{Hopf,Cole} mapping it to the linear heat equation. 

Many papers about nonclassical symmetries of classes of Burgers' systems, as well as symmetries of systems of coupled Burgers-heat equations are available in the literature (see, for instance, \cite{Mansfield1999,Arrigo-Hickling,Arrigo-Ekrut,Cherniha-Serov,Webb}). 

It is known \cite{Olver-Vorobev} that nonclassical symmetries of the classical Burgers' equation \eqref{classicalburgers} express in terms of three functions $a(t,x)$, $b(t,x)$ and $c(t,x)$ satisfying a coupled system of Burgers-like equations, say
\begin{equation}
\label{threeburgers}
\begin{aligned}
&a_{,t}+aa_{,x}-a_{,xx}+b_{,x}=0,\\
&b_{,t}+ba_{,x}-b_{,xx}+c_{,x}=0,\\
&c_{,t}+ca_{,x}-c_{,xx}=0.
\end{aligned}
\end{equation}
Hereafter, we first prove that some conditional symmetries of the system \eqref{threeburgers} are expressed in terms of five functions satisfying a system of coupled Burgers-like equations with a similar structure. Using the latter system as a starting point and iterating the procedure, an infinite hierarchy of systems made of an odd number of coupled Burgers-like equations seems to arise. Moreover, we also prove that, starting from the pair of  Burgers-like equations \eqref{coupleburgers}, a similar hierarchy of systems made of an even number of coupled Burgers-like equations arises. 

The main result of this paper (Theorem~\ref{thm:main}) consists in proving that the two hierarchies may be unified; in fact, for every non-vanishing $m\in\mathbb{N}$, a system of $m$ coupled Burgers-like equations (the $k$th element of the hierarchy, where  $k = \lceil m/2 \rceil$), namely
\begin{equation}
\boldsymbol\Delta_m\equiv\left\{
\begin{aligned}
&u^{(k)}_{\alpha,t}+u^{(k)}_\alpha u^{(k)}_{1,x}-u^{(k)}_{\alpha,xx}+u^{(k)}_{\alpha+1,x}=0,\\
&u^{(k)}_{m,t}+u^{(k)}_m u^{(k)}_{1,x}-u^{(k)}_{m,xx}=0,
\end{aligned}
\right.
\end{equation}
with $\alpha=1,\ldots,m-1$, admits $Q$-conditional symmetries that are
expressed in terms of the solutions of a similar system of coupled Burgers-like equations involving two more unknown functions, \emph{i.e.},
\begin{equation}
\boldsymbol{\Delta}_{m+2}\equiv
\left\{
\begin{aligned}
&u^{(k+1)}_{\alpha,t}+u^{(k+1)}_\alpha u^{(k+1)}_{1,x}-u^{(k+1)}_{\alpha,xx}+u^{(k+1)}_{\alpha+1,x}=0,\\
&u^{(k+1)}_{m+2,t}+u^{(k+1)}_{m+2}u^{(k+1)}_{1,x}-u^{(k+1)}_{m+2,xx}=0,
\end{aligned}
\right.
\end{equation}
with $\alpha=1,\ldots,m+1$.

Before proving the general result, in Section~\ref{sec:3}, we analyze in detail the $Q$-conditional symmetries of the first elements of the hierarchies made of an odd or even number of equations. These computations allow us to guess the structure of the vector fields of nonclassical symmetries for the generic element of the hierarchy and so state the main theorem.
Moreover, writing a generic element of this hierarchy as a matrix Burgers' equation \cite{Mansfield1999,Arrigo-Hickling}, the existence of the matrix Hopf-Cole transformation \cite{Levi1983} allows for its linearization and the determination of its solution in terms of the solutions of linear heat equations.
Finally, we show that all the elements of this infinite hierarchy share the property of admitting a five-dimensional Lie algebra of point symmetries; these Lie algebras, even if realized in terms of vector fields on manifolds with different dimensionality, are all isomorphic.

The idea of iterating the nonclassical method for evolution equations, and deducing the so-called heir equations, dates back to 1994 \cite{Nucci1994} (see also  \cite{Allassia-Nucci} and the recent review \cite{Nucci-25years}), and has been proved useful for determining solutions of starting equations. Anyway, we remark that the approach followed in this paper is rather different, as will be shown below; moreover, differently from \cite{Nucci1994}, we will not restrict ourselves to consider the no-go case \cite{Arrigo-Ekrut}.

The plan of the paper is as follows. In Section~\ref{sec:2}, for the reader's convenience, a brief sketch of the approach to conditional symmetries is given. In Section~\ref{sec:3}, the conditional symmetries admitted by some special instances of systems of Burgers-like equations are investigated. 
The results here derived will serve as a guide for stating the general theorem proved in Section~\ref{sec:4}. Moreover,  every element of the hierarchy can be written as a matrix Burgers' equation and linearized by means of the matrix Hopf-Cole transform; it is also shown that each member of the hierarchy possesses a five--dimensional Lie algebra of classical point symmetries. Finally, Section~\ref{sec:5} contains our conclusions.

\section{Theoretical preliminaries}
\label{sec:2}
In this Section, also to fix the notation, we briefly review the used approach to conditional symmetries.

Let us consider an $r$th order differential equation, say
\begin{equation}
\label{de}
 \Delta\left(x_i,u_\alpha,u_{\alpha,i},\ldots,u_{\alpha,i_1,\ldots,i_r}\right)=0,
\end{equation}
where $x_i$ $(i=1,\ldots,n)$ are the independent variables, $u_\alpha$ $(\alpha=1,\ldots,m)$ the dependent variables, and $\displaystyle u_{\alpha,i_1,\ldots,i_k}=\frac{\partial^k u_\alpha}{\partial x_{i_1}\ldots\partial x_{i_k}}$ $(k=1,\ldots,r)$.

A Lie point symmetry of \eqref{de} is characterized by the infinitesimal operator
\begin{equation}
\label{op}
\Xi=\sum_{i=1}^n\xi_i(x_j,u_\beta)\frac{\partial}{\partial x_i}
+\sum_{\alpha=1}^m\eta_\alpha(x_j,u_\beta)\frac{\partial}{\partial u_\alpha}
\end{equation}
such that
\begin{equation}
\label{invcond}
\left.\Xi^{(r)}\left(\Delta\right)\right|_{\Delta=0}=0,
\end{equation}
where $\Xi^{(r)}$ is the $r$th prolongation of \eqref{op} 
\cite{Ovsiannikov,Olver1,Ibragimov:CRC}. Condition \eqref{invcond} leads to 
a system of linear partial differential equations  (\emph{determining equations}) whose integration provides the \emph{infinitesimals} $\xi_i$ and $\eta_\alpha$. Invariant solutions corresponding to a given Lie point symmetry are found by solving 
the invariant surface conditions
\begin{equation}
\label{Qconstraint}
Q_\alpha\equiv \sum_{i=1}^n \xi_i(x_j,u_\beta) u_{\alpha,i}-
{\eta_\alpha}(x_j,u_\beta)=0,\qquad \alpha=1,\ldots,m,
\end{equation}
and inserting their solutions in \eqref{de}.

In 1969, Bluman and Cole \cite{Bluman-Cole-1969} introduced a generalization of classical Lie symmetries, and applied their method (called \emph{nonclassical}) to the linear heat equation. The basic idea was that of imposing the invariance to a system made by the differential equation at hand, the invariance surface condition together with the differential consequences of the latter. This method requires to solve a set of \emph{nonlinear} determining equations whose general integration is usually difficult. Nonclassical symmetries are now part of conditional symmetries, \emph{i.e.}, symmetries of differential equations where some additional differential conditions are imposed to restrict the set of solutions. This method revealed useful in many applied problems modeled by differential equations (for instance, reaction-diffusion equations \cite{Cherniha-JMAA2007,Cherniha-JPA2010,Cherniha-Davydovych}) possessing very few Lie point symmetries; consequently, 
more rich reductions leading to wide classes of exact solutions are possible. 

The nonclassical symmetries introduced by Bluman and Cole are now referred to as  
$Q$-conditional symmetries \cite{Cherniha-Davydovych}. In such a case, $Q$-conditional symmetries are expressed by vector fields $\Xi$ such that
\begin{equation}
\label{invcond-gen}
\left.\Xi^{(r)}(\Delta)\right|_{\mathcal{M}}=0,
\end{equation}
where $\mathcal{M}$ is the manifold of the jet space defined by 
\begin{equation}
\Delta=0, \qquad Q_\alpha=0, \qquad \frac{D}{D x_{j_1}}
\frac{D}{D x_{j_2}}\cdots \frac{D}{Dx_{j_k}}Q_\alpha=0,
\end{equation}
with $1\le j_1,j_2,\ldots,j_k \le n$, $1\le k\le r-1$, and $\alpha=1,\ldots,m$, along with the Lie derivative
\begin{equation}
\begin{aligned}
&\frac{D}{D x_k} = \frac{\partial}{\partial x_k} + u_{\alpha,k}\frac{\partial}{\partial u_\alpha}+u_{\alpha,ik}\frac{\partial}{\partial u_{\alpha,i}}+u_{\alpha,ijk}\frac{\partial}{\partial u_{\alpha,ij}}+\cdots,
\end{aligned}
\end{equation}
where the Einstein convention on sums over repeated indices has been used.

Trivially, a (classical) Lie symmetry is a $Q$-conditional symmetry. However, differently from Lie symmetries, all possible conditional symmetries of a differential equation form a set which is neither a Lie algebra nor a linear space in the general case. Furthermore, if the vector field  of a 
$Q$-conditional symmetry is multiplied by an arbitrary nonvanishing smooth function of dependent and independent variables, we have still a $Q$-conditional symmetry.

In the following, we will be concerned with second order partial differential equations ruling the evolution of $m$ unknown functions depending on $t$ and $x$, and consider $Q$-conditional symmetries corresponding to the vector field
\begin{equation}
\Xi = \frac{\partial}{\partial t} + \xi(t,x,u_\beta)\frac{\partial}{\partial x} + \sum_{\alpha=1}^m\eta_\alpha(t,x,u_\beta) \frac{\partial}{\partial u_\alpha}.
\end{equation}

Below it will be shown that, starting with the classical Burgers' equation or with a special pair of coupled Burgers-like equations,  and looking for $Q$-conditional symmetries, an infinite hierarchy of coupled systems of Burgers-like equations is recovered.
 
\section{Conditional symmetries of Burgers-like\\ equations}
\label{sec:3}
In this Section, we start considering the classical Burger's equation \cite{Crighton}. In \cite{Olver-Vorobev},  it was proved that the $Q$-conditional symmetries of Burgers' equation are expressed in terms of three functions that are solutions of a system of coupled Burgers-like equations. In what follows, we prove that the latter system of coupled Burgers-like equations admits $Q$-conditional symmetries expressed in terms of five functions satisfying a new system of coupled Burgers-like equations. This process can be repeatedly used, and a hierarchy of systems involving an odd number of unknowns arises. Moreover, we prove also that, starting with a pair of coupled Burgers-like equations,  another hierarchy of systems involving an even number of coupled Burgers-like equations is generated.
 
\subsection{Hierarchy originating from  Burgers' equation}
\label{subsec:3.1}
Let us consider the Burgers' equation
\begin{equation}
\label{eqn:Burgers}
\Delta_1\equiv u^{(1)}_{,t} + u^{(1)}u^{(1)}_{,x} - u^{(1)}_{,xx} = 0
\end{equation}
for the unknown $u^{(1)}(t,x)$, and take the vector field
\begin{equation}
\label{vf-conditional}
\Xi = \frac{\partial}{\partial t} + \xi(t,x,u^{(1)})\frac{\partial}{\partial x} + \eta(t,x,u^{(1)}) \frac{\partial}{\partial u^{(1)}}.
\end{equation}
In order to compute the $Q$-conditional symmetries of \eqref{eqn:Burgers} associated to \eqref{vf-conditional}, let us define the manifold $\mathcal{M}_1$ as
\begin{equation}
\left\{
\begin{aligned}
&\Delta_1=0 ,\\
&Q_1\equiv \Xi\left(u^{(1)}-u^{(1)}(t,x)\right)=0,\\
&\frac{D Q_1}{Dt} = \frac{DQ_1}{Dx}=0,
\end{aligned}
\right.
\end{equation}
whence the conditional symmetries are found by requiring 
\[
\left.\Xi^{(2)}(\Delta_1)\right|_{\mathcal{M}_1}=0.
\]
The latter provides the following polynomial of third degree in the derivative $u^{(1)}_{,x}$
\begin{equation*}
\label{burg1}
\begin{aligned}
&\frac{\partial^2 \xi}{\partial u^{{(1)}2}}\left(u^{(1)}_{,x}\right)^3 +\left(2\frac{\partial^2 \xi}{\partial x\partial u^{(1)}}-\frac{\partial^2 \eta}{\partial u^{{(1)}2}} + 2\frac{\partial \xi}{\partial u^{(1)}}u^{(1)} -2\xi \frac{\partial \xi}{\partial u^{(1)}}\right) \left(u^{(1)}_{,x}\right)^2\\
&+\left(\frac{\partial^2 \xi}{\partial x^2}-2\frac{\partial^2 \eta}{\partial x\partial u^{(1)}} - \frac{\partial \xi}{\partial t} + \frac{\partial \xi}{\partial x}u^{(1)}- 2\xi\frac{\partial \xi}{\partial x}+2 \frac{\partial \xi}{\partial u^{(1)}} \eta + \eta\right)u^{(1)}_{,x}\\
&- \frac{\partial^2 \eta}{\partial x^2}+ 2 \frac{\partial \xi}{\partial x} \eta+\frac{\partial \eta}{\partial t}  + \frac{\partial \eta}{\partial x} u^{(1)} = 0.
\end{aligned}
\end{equation*}
Annihilating the coefficients of this polynomial, after simple algebra, we get
\begin{equation}
\label{eqn:xiBurgers}
\begin{aligned}
&\xi = \kappa u^{(1)} +\frac{1}{2}u^{(2)}_1,\\
&\eta= \frac{\kappa(1-\kappa)}{3} \left(u^{(1)}\right)^3 - \frac{\kappa}{2} u^{(2)}_1\left(u^{(1)}\right)^2 + \frac{1}{4}u^{(1)}u^{(2)}_2 + \frac{1}{4} u^{(2)}_3,
\end{aligned}
\end{equation}
where $\kappa$ is a constant such that $\kappa(\kappa-1)(2\kappa+1) = 0$, whereas $u^{(2)}_1(t,x)$, $u^{(2)}_2(t,x)$ and 
$u^{(2)}_3(t,x)$ are functions depending on the indicated arguments. 

Three cases must be distinguished,  say $\kappa= 0$, $\kappa = 1$ and $\kappa = -1/2$, the latter being the most interesting one. In such a case, the functions $u^{(2)}_1(t,x)$, $u^{(2)}_2(t,x)$ and 
$u^{(2)}_3(t,x)$ satisfy the system
\begin{equation}
\label{eqn:sysBurgersLike3}
\boldsymbol{\Delta}_3\equiv
\left\{
\begin{aligned}
&u^{(2)}_{1,t} + u^{(2)}_1u^{(2)}_{1,x} - u^{(2)}_{1,xx} + u^{(2)}_{2,x} =0,\\
&u^{(2)}_{2,t} + u^{(2)}_2u^{(2)}_{1,x} - u^{(2)}_{2,xx} + u^{(2)}_{3,x} =0,\\
&u^{(2)}_{3,t} + u^{(2)}_3u^{(2)}_{1,x} - u^{(2)}_{3,xx}=0.
\end{aligned}
\right.
\end{equation}
As already remarked, this result has been obtained in \cite{Olver-Vorobev}.

Then, it can be interesting to explore the $Q$-conditional symmetries admitted by the system (\ref{eqn:sysBurgersLike3}). As a result, the following Proposition is proved.

\begin{prop}
\label{prop1}
The vector field 
\begin{equation}
\label{xi-3}
\Xi = \frac{\partial}{\partial t} + \xi(t,x,u^{(2)}_\beta)\frac{\partial}{\partial x} +\sum_{\alpha=1}^3 \eta_{\alpha}(t,x,u^{(2)}_\beta) \frac{\partial}{\partial u^{(2)}_\alpha}
\end{equation}
gives a $Q$-conditional symmetry of the system \eqref{eqn:sysBurgersLike3}
provided that
\begin{equation}
\label{eqn:infgen3}
\begin{aligned}
&\xi=\frac{1}{2}\left(-u^{(2)}_1+u_1^{(3)}\right),\\
&\eta_{1}=\frac{1}{4}\left(-\left(u_1^{(2)}\right)^{3}-2u^{(2)}_1u^{(2)}_2+u^{(3)}_1\left(u^{(2)}_1\right)^{2}+u^{(3)}_2u^{(2)}_1+u^{(3)}_1u^{(2)}_2\right.\\
&\left.\phantom{\frac{1}{}}-u^{(2)}_3+u^{(3)}_3\right),\\
&\eta_{2}=\frac{1}{4}\left(-\left(u^{(2)}_1\right)^{2}u^{(2)}_2-u^{(2)}_1u^{(2)}_3-\left(u^{(2)}_2\right)^{2}+u^{(3)}_1u^{(2)}_1u^{(2)}_2+u^{(3)}_2u^{(2)}_2\right.\\
&\left.\phantom{\frac{1}{}}+u^{(3)}_1u^{(2)}_3+u^{(3)}_4\right),\\
&\eta_{3}=\frac{1}{4}\left(-\left(u^{(2)}_1\right)^{2}u^{(2)}_3-u^{(2)}_2u^{(2)}_3+u^{(3)}_1u^{(2)}_1u^{(2)}_3+u^{(3)}_2u^{(2)}_3+u^{(3)}_5\right),
\end{aligned}
\end{equation}
where the functions $u^{(3)}_\alpha(t,x)$ ($\alpha=1,\ldots,5$) satisfy the constraints
\begin{equation}
\label{eqn:sysBurgersLike5}
\boldsymbol{\Delta}_5\equiv
\left\{
\begin{aligned}
&u^{(3)}_{1,t} + u^{(3)}_1u^{(3)}_{1,x} - u^{(3)}_{1,xx} + u^{(3)}_{2,x} = 0 ,\\
&u^{(3)}_{2,t} + u^{(3)}_2u^{(3)}_{1,x} - u^{(3)}_{2,xx} + u^{(3)}_{3,x} = 0 ,\\
&u^{(3)}_{3,t} + u^{(3)}_3u^{(3)}_{1,x} - u^{(3)}_{3,xx} + u^{(3)}_{4,x} = 0 ,\\
&u^{(3)}_{4,t} + u^{(3)}_4u^{(3)}_{1,x} - u^{(3)}_{4,xx} + u^{(3)}_{5,x} = 0 ,\\
&u^{(3)}_{5,t} + u^{(3)}_5u^{(3)}_{1,x} - u^{(3)}_{5,xx} = 0.
\end{aligned}
\right.
\end{equation}
\end{prop}
\proof
The proof immediately follows by requiring 
\begin{equation}
\left.\Xi^{(2)}(\boldsymbol{\Delta}_3)\right|_{\mathcal{M}_3} = \mathbf{0},
\end{equation}
where $\mathcal{M}_3$ is the manifold of the jet space defined by the system \eqref{eqn:sysBurgersLike3} together with the invariant surface conditions and their differential consequences, say
\begin{equation}
\left\{
\begin{aligned}
&\boldsymbol{\Delta}_3=\mathbf{0},\\
&Q_\alpha\equiv \Xi\left(u^{(2)}_\alpha- u^{(2)}_{\alpha}(t,x)\right)=0,\qquad \alpha=1,2,3,\\
&\frac{DQ_\alpha}{Dt} = \frac{DQ_\alpha}{Dx}=0.
\end{aligned}
\right.
\end{equation}
The lengthy computations can be done by using the program \texttt{ReLie} \cite{Oliveri:relie} written in the Computer Algebra System \emph{Reduce} \cite{Reduce}. As a result, we obtain three polynomials of third degree in the derivatives $u^{(2)}_{\alpha,x}$. We immediately obtain 
\begin{equation}
\xi = \kappa u^{(2)}_1 +\frac{1}{2}u^{(3)}_1,
\end{equation} 
$u^{(3)}_1(t,x) $ being a function of the indicated arguments, and $\kappa$ is a constant that has to satisfy the condition 
\begin{equation}
\kappa(2\kappa + 1) = 0.
\end{equation}
The most interesting case again corresponds to the choice  $\kappa = -1/2$. After straightforward though tedious  computations, all the determining equations can be solved, and the vector field \eqref{xi-3} assumes the form (\ref{eqn:infgen3}), along with the functions $u^{(3)}_\alpha(t,x)$ ($\alpha=1,\ldots,5$) that satisfy the system of differential equations \eqref{eqn:sysBurgersLike5}.
\endproof

\begin{remark}
We note that the system (\ref{eqn:sysBurgersLike5}) has the same structure as the system (\ref{eqn:sysBurgersLike3}), even if it involves two more unknowns.
\end{remark}

Nothing prevents us to repeat the procedure looking for $Q$-conditional symmetries of system \eqref{eqn:sysBurgersLike5}. The result we obtain is stated with the following Proposition.
\begin{prop}
There exist $Q$-conditional symmetries of  \eqref{eqn:sysBurgersLike5} in correspondence to the vector field
\begin{equation}
\label{xi-5}
\Xi = \frac{\partial}{\partial t} + \xi(t,x,u^{(3)}_\beta)\frac{\partial}{\partial x} +\sum_{\alpha=1}^5 \eta_{\alpha}(t,x,u^{(3)}_\beta) \frac{\partial}{\partial u^{(3)}_\alpha},
\end{equation}
where
\begin{equation}
\begin{aligned}
&\xi=\frac{1}{2}\left(-u^{(3)}_1+u^{(4)}_1\right),\\
&\eta_{1}=\frac{1}{4}\left(-\left(u^{(3)}_1\right)^{3}-2u^{(3)}_1u^{(3)}_2+u^{(4)}_1\left(u^{(3)}_1\right)^{2}+u^{(4)}_2u^{(3)}_1\right.\\
&\left.\phantom{\frac{1}{}}+u^{(4)}_1u^{(3)}_2-u^{(3)}_3+u^{(4)}_3\right),\\
&\eta_{2}=\frac{1}{4}\left(-\left(u^{(3)}_1\right)^{2}u^{(3)}_2-u^{(3)}_1u^{(3)}_3-\left(u^{(3)}_2\right)^{2}+u^{(4)}_1u^{(3)}_1u^{(3)}_2+u^{(4)}_2u^{(3)}_2\right.\\
&\left.\phantom{\frac{1}{}}+u^{(4)}_1u^{(3)}_3-u^{(3)}_4+u^{(4)}_4\right),\\
&\eta_{3}=\frac{1}{4}\left(-\left(u^{(3)}_1\right)^{2}u^{(3)}_3-u^{(3)}_1u^{(3)}_4-u^{(3)}_2u^{(3)}_3+u^{(4)}_1u^{(3)}_1u^{(3)}_3+u^{(4)}_2u^{(3)}_3\right.\\
&\left.\phantom{\frac{1}{}}+u^{(4)}_1u^{(3)}_4-u^{(3)}_5+u^{(4)}_5\right),\\
&\eta_{4}=\frac{1}{4}\left(-\left(u_1^{(3)}\right)^{2}u^{(3)}_4-u^{(3)}_1u^{(3)}_5-u^{(3)}_2u^{(3)}_4+u^{(4)}_1u^{(3)}_1u^{(3)}_4+u^{(4)}_2u^{(3)}_4\right.\\
&\left.\phantom{\frac{1}{}}+u^{(4)}_1u^{(3)}_5+u^{(4)}_6\right),\\
&\eta_{5}=\frac{1}{4}\left(-\left(u^{(3)}_1\right)^{2}u^{(3)}_5-u^{(3)}_2u^{(3)}_5+u^{(4)}_1u^{(3)}_1u^{(3)}_5+u^{(4)}_2u^{(3)}_5+u^{(4)}_7\right),
\end{aligned}
\end{equation}
and the functions $u^{(4)}_\alpha(t,x)$ $(\alpha=1,\ldots,7)$ satisfy the system
\begin{equation}
\boldsymbol{\Delta}_7 \equiv 
\left\{
\begin{aligned}
&u^{(4)}_{1,t} + u^{(4)}_1u^{(4)}_{1,x} - u^{(4)}_{1,xx} + u^{(4)}_{2,x} = 0,\\
&u^{(4)}_{2,t} + u^{(4)}_2u^{(4)}_{1,x} - u^{(4)}_{2,xx} + u^{(4)}_{3,x} = 0,\\
&u^{(4)}_{3,t} + u^{(4)}_3u^{(4)}_{1,x} - u^{(4)}_{3,xx} + u^{(4)}_{4,x} = 0,\\
&u^{(4)}_{4,t} + u^{(4)}_4u^{(4)}_{1,x} - u^{(4)}_{4,xx} + u^{(4)}_{5,x} = 0,\\
&u^{(4)}_{5,t} + u^{(4)}_5u^{(4)}_{1,x} - u^{(4)}_{5,xx} + u^{(4)}_{6,x} = 0,\\
&u^{(4)}_{6,t} + u^{(4)}_6u^{(4)}_{1,x} - u^{(4)}_{6,xx} + u^{(4)}_{7,x} = 0,\\
&u^{(4)}_{7,t} + u^{(4)}_7u^{(4)}_{1,x} - u^{(4)}_{7,xx} = 0.
\end{aligned}
\right.
\end{equation}
\end{prop}
\proof
The proof requires only straightforward though lengthy computations. Also in this case the Reduce program \texttt{ReLie} has been used.
\endproof

The results heretofore obtained can be summarized as follows:
\begin{itemize} 
\item there are $Q$-conditional symmetries of the Burgers' equation expressed in terms of three functions representing arbitrary solutions of the system $\boldsymbol{\Delta}_3$ made of three coupled Burgers-like equations;
\item there are $Q$-conditional symmetries of $\boldsymbol{\Delta}_3$ expressed in terms of five functions representing arbitrary solutions of the system $\boldsymbol{\Delta}_5$ made of five coupled Burgers-like equations; 
\item  there are $Q$-conditional symmetries of $\boldsymbol{\Delta}_5$ expressed in terms of seven functions representing arbitrary solutions of the system $\boldsymbol{\Delta}_7$ made of seven coupled Burgers-like equations.
\end{itemize}

It seems natural to conjecture that repeatedly searching for $Q$-conditional symmetries, and starting from the classical Burgers' equation, a hierarchy of systems made of an odd number of Burgers-like equations may arise.
  
In the next Subsection, we shall consider the case of a coupled system made of an even number of Burgers-like equations. In particular, the starting point will be the system of two Burgers-like equations whose structure is deduced from \eqref{eqn:sysBurgersLike3} where we set $u^{(2)}_3\equiv 0$.

\subsection{Hierarchy originating from a pair of coupled\\ Burgers-like equations}
\label{subsec:3.2}
Let us consider the following system made of two coupled Burgers-like equations
\begin{equation}
\label{eqn:BurgersLike2}
\boldsymbol\Delta_2\equiv\left\{
\begin{aligned}
&u^{(1)}_{1,t} + u^{(1)}_1u^{(1)}_{1,x} - u^{(1)}_{1,xx} + u^{(1)}_{2,x} =  0 ,  \\
&u^{(1)}_{2,t} + u^{(1)}_2u^{(1)}_{1,x} - u^{(1)}_{2,xx} =  0.
\end{aligned}
\right.
\end{equation}

By looking for $Q$-conditional symmetries of \eqref{eqn:BurgersLike2} in correspondence to the vector field
\begin{equation}
\label{xi-2}
\Xi = \frac{\partial}{\partial t} + \xi(t,x,u^{(1)}_\beta)\frac{\partial}{\partial x} + \eta_{1}(t,x,u^{(1)}_\beta) \frac{\partial}{\partial u^{(1)}_1} + \eta_{2}(t,x,u^{(1)}_\beta) \frac{\partial}{\partial u^{(1)}_2},
\end{equation}
and requiring that
\begin{equation}
\Xi^{(2)}(\mathbf{\Delta}_2)\Big\rvert_{\mathcal{M}_2} = \mathbf{0},
\end{equation}
where $\mathcal{M}_2$ is the manifold of the jet space defined by the system \eqref{eqn:BurgersLike2} together with the invariant surface conditions and their differential consequences,  say
\begin{equation}
\left\{
\begin{aligned}
&\boldsymbol{\Delta}_2=\mathbf{0},\\
&Q_\alpha\equiv \Xi\left(u^{(1)}_\alpha- u^{(1)}_{\alpha}(t,x)\right)=0,\qquad \alpha=1,2,\\
&\frac{DQ_\alpha}{Dt} = \frac{DQ_\alpha}{Dx}=0,
\end{aligned}
\right.
\end{equation}
we obtain the following invariance conditions:
\begin{align*}
&\frac{\partial^2\xi}{\partial u^{(1)2}_1}\left(u^{(1)}_{1,x}\right)^3+ 2\frac{\partial^2\xi}{\partial u^{(1)}_1 \partial u^{(1)}_2}\left(u^{(1)}_{1,x}\right)^2u^{(1)}_{2,x}+ \frac{\partial^2\xi}{\partial u_2^{(1)2}}u^{(1)}_{1,x}\left(u^{(1)}_{2,x}\right)^2 \allowdisplaybreaks\notag\\
&+\left(2\frac{\partial^2\xi}{\partial x \partial u^{(1)}_1}- \frac{\partial^2\eta_{1}}{\partial u_1^{(1)2}}+ 2\frac{\partial\xi}{\partial u^{(1)}_1}u^{(1)}_1- 2\xi\frac{\partial\xi}{\partial u^{(1)}_1}+ \frac{\partial\xi}{\partial u^{(1)}_2}u^{(1)}_2 \right)\left(u^{(1)}_{1,x}\right)^2\allowdisplaybreaks\notag\\
&+ \left(2\frac{\partial^2\xi}{\partial x \partial u^{(1)}_2}- 2\frac{\partial^2\eta_{1}}{\partial u^{(1)}_1 \partial u^{(1)}_2}+ 2\frac{\partial\xi}{\partial u^{(1)}_1}- 2\xi\frac{\partial\xi}{\partial u^{(1)}_2}+ \frac{\partial\xi}{\partial u^{(1)}_2}u^{(1)}_1\right)u^{(1)}_{1,x}u^{(1)}_{2,x}\allowdisplaybreaks\notag\\
&+\left( \frac{\partial\xi}{\partial u^{(1)}_2}- \frac{\partial^2\eta_{1}}{\partial u^{(1)2}_2}\right)\left(u^{(1)}_{2,x}\right)^2+ \left(\frac{\partial^2\xi}{\partial x^2}- 2\frac{\partial^2\eta_{1}}{\partial x\partial u^{(1)}_1}- \frac{\partial\xi}{\partial t}+ \frac{\partial\xi}{\partial x}u^{(1)}_1\right.\allowdisplaybreaks\notag\\
&\left. - 2\xi\frac{\partial\xi}{\partial x}+2\eta_{1}\frac{\partial\xi}{\partial u^{(1)}_1}- \frac{\partial\eta_{1}}{\partial u^{(1)}_2}u^{(1)}_{2}+ \frac{\partial\eta_{2}}{\partial u^{(1)}_1}+ \eta_{1}\right)u^{(1)}_{1,x}\allowdisplaybreaks\notag\\
&+\left(- 2\frac{\partial\eta_{1}}{\partial x\partial u^{(1)}_2}+\frac{\partial\xi}{\partial x}+ 2\eta_{1}\frac{\partial\xi}{\partial u^{(1)}_2}- \frac{\partial\eta_{1}}{\partial u^{(1)}_1} + \frac{\partial\eta_{1}}{\partial u^{(1)}_2}u^{(1)}_1+ \frac{\partial\eta_{2}}{\partial u^{(1)}_2}\right)u^{(1)}_{2,x}\allowdisplaybreaks\notag\\
&- \frac{\partial^2\eta_{1}}{\partial x^2}+2\eta_{1}\frac{\partial\xi}{\partial x}+ \frac{\partial\eta_{1}}{\partial t}+ \frac{\partial\eta_{1}}{\partial x}u^{(1)}_1 + \frac{\partial\eta_{2}}{\partial x} = 0, \allowdisplaybreaks\notag\\
&\frac{\partial^2\xi}{\partial u^{(1)2}_1}\left(u^{(1)}_{1,x}\right)^2u^{(1)}_{2,x}+ 2\frac{\partial^2\xi}{\partial u^{(1)}_1 \partial u^{(1)}_2}u^{(1)}_{1,x}\left(u^{(1)}_{2,x}\right)^2\\
&+ \frac{\partial^2\xi}{\partial u^{(1)2}_2}\left(u^{(1)}_{2,x}\right)^3+ \left(\frac{\partial\xi}{\partial u^{(1)}_1}u^{(1)}_2- \frac{\partial^2\eta_{2}}{\partial u^{(1)2}_1}\right)\left(u^{(1)}_{1,x}\right)^2\allowdisplaybreaks\notag\\
&+\left(2\frac{\partial^2\xi}{\partial x\partial u^{(1)}_1}- 2\frac{\partial^2\eta_{2}}{\partial u^{(1)}_1 \partial u^{(1)}_2}+ \frac{\partial\xi}{\partial u^{(1)}_1}u^{(1)}_1- 2\xi\frac{\partial\xi}{\partial u^{(1)}_1}+ 2\frac{\partial\xi}{\partial u^{(1)}_2}u^{(1)}_{2}\right)u^{(1)}_{1,x}u^{(1)}_{2,x}\allowdisplaybreaks\notag\\
&+\left(2\frac{\partial^2\xi}{\partial x\partial u^{(1)}_2}- \frac{\partial^2\eta_{2}}{\partial u^{(1)2}_2}+ \frac{\partial\xi}{\partial u^{(1)}_1}- 2\xi\frac{\partial\xi}{\partial u^{(1)}_2}\right)\left(u^{(1)}_{2,x}\right)^2+\left(\frac{\partial\xi}{\partial x}u^{(1)}_2- 2\frac{\partial^2\eta_{2}}{\partial x\partial u^{(1)}_1}\right.\allowdisplaybreaks\notag\\
&\left.+ 2\eta_{2}\frac{\partial\xi}{\partial u^{(1)}_1}+\frac{\partial\eta_{1}}{\partial u^{(1)}_1}u^{(1)}_{2}-\frac{\partial\eta_{2}}{\partial u^{(1)}_1}u^{(1)}_1- \frac{\partial\eta_{2}}{\partial u^{(1)}_2}u^{(1)}_{2}+ \eta_{2}\right)u^{(1)}_{1,x}\allowdisplaybreaks\notag\\
&+\left(\frac{\partial^2\xi}{\partial x^2}- 2\frac{\partial^2\eta_{2}}{	\partial x\partial u^{(1)}_2}- \frac{\partial\xi}{\partial t}- 2\xi\frac{\partial\xi}{\partial x}+ 2\eta_{2}\frac{\partial\xi}{\partial u^{(1)}_2}+\frac{\partial\eta_{1}}{\partial u^{(1)}_2}u^{(1)}_2- \frac{\partial\eta_{2}}{\partial u^{(1)}_1} \right)u^{(1)}_{2,x}\allowdisplaybreaks\notag\\
& - \frac{\partial^2\eta_{2}}{\partial x^2} + 2\eta_{2}\frac{\partial\xi}{\partial x}+ \frac{\partial\eta_{2}}{\partial t}+ \frac{\partial\eta_{1}}{\partial x}u^{(1)}_2  = 0;
\end{align*}
the latter are polynomials of third degree in the derivatives $u^{(1)}_{1,x}$ and $u^{(1)}_{2,x}$, whose coefficients must be vanishing. After simple algebra, we get
\begin{equation}
\xi = \kappa u^{(1)}_1 +\frac{1}{2} u^{(2)}_1,
\end{equation}
where $u^{(2)}_1(t,x)$ is a function of the indicated arguments, and $\kappa$ is a constant that has to satisfy the constraint $\kappa(2\kappa+1)=0$. Again, looking for the $Q$-conditional symmetries of system \eqref{eqn:BurgersLike2}, we choose $\kappa=-1/2$. Thence, integrating the determining equations, we obtain the $Q$-conditional symmetries characterized by the vector field \eqref{xi-2}, with
\begin{equation}
\label{eqn:infgen2}
\begin{aligned}
&\xi=\frac{1}{2}\left(-u^{(1)}_1+u^{(2)}_1\right),\\
&\eta_{1}=\frac{1}{4}\left(-\left(u^{(1)}_1\right)^{3}-2u^{(1)}_1u^{(1)}_2+u^{(2)}_1\left(u^{(1)}_1\right)^{2}+u^{(2)}_2u^{(1)}_1\right.\\
&\left.\phantom{\frac{1}{}}+u^{(2)}_1u^{(1)}_2+u^{(2)}_3\right),\\
&\eta_{2}=\frac{1}{4}\left(-\left(u^{(1)}_1\right)^{2}u^{(1)}_2-\left(u^{(1)}_2\right)^{2}+u^{(2)}_1u^{(1)}_1u^{(1)}_2+u^{(2)}_2u^{(1)}_2+u^{(2)}_4\right),
\end{aligned}
\end{equation}
and the functions $u^{(2)}_\alpha(t,x)$ $(\alpha=1,\ldots,4)$ satisfying the constraints
\begin{equation}
\label{eqn:sysBurgersLike4}
\boldsymbol{\Delta}_4\equiv
\left\{
\begin{aligned}
&u^{(2)}_{1,t} + u^{(2)}_1u^{(2)}_{1,x} - u^{(2)}_{1,xx} + u^{(2)}_{2,x} = 0,\\
&u^{(2)}_{2,t} + u^{(2)}_2u^{(2)}_{1,x} - u^{(2)}_{2,xx} + u^{(2)}_{3,x} = 0,\\
&u^{(2)}_{3,t} + u^{(2)}_3u^{(2)}_{1,x} - u^{(2)}_{3,xx} + u^{(2)}_{4,x} = 0,\\
&u^{(2)}_{4,t} + u^{(2)}_4u^{(2)}_{1,x} - u^{(2)}_{4,xx} = 0.
\end{aligned}
\right.
\end{equation}
Repeating the same algorithm for the latter system of four coupled Burgers-like equations, the admitted $Q$-conditional symmetries are expressed in terms of six arbitrary functions depending on $t$ and $x$. We write this result in the following Proposition.
\begin{prop}
The system \eqref{eqn:sysBurgersLike4}
admits the vector field $\Xi$ of the $Q$-conditional symmetries, say
\begin{equation}
\label{xi-4}
\Xi = \frac{\partial}{\partial t} + \xi(t,x,u^{(2)}_\beta)\frac{\partial}{\partial x} +\sum_{\alpha=1}^4 \eta_{\alpha}(t,x,u^{(2)}_\beta) \frac{\partial}{\partial u^{(2)}_\alpha},
\end{equation}
where
\begin{align}
&\xi=\frac{1}{2}\left(-u^{(2)}_1+u^{(3)}_1\right),\allowdisplaybreaks\notag\\
&\eta_{1}=\frac{1}{4}\left(-\left(u^{(2)}_1\right)^{3}-2u^{(2)}_1u^{(2)}_2+u^{(3)}_1\left(u^{(2)}_1\right)^{2}+u^{(3)}_2u^{(2)}_1\right.\allowdisplaybreaks\notag\\
&\left.\phantom{\frac{1}{}}+u^{(3)}_1u^{(2)}_2-u^{(2)}_3+u^{(3)}_3\right),\allowdisplaybreaks\notag\\
& \eta_{2}=\frac{1}{4}\left(-\left(u^{(2)}_1\right)^{2}u^{(2)}_2-u^{(2)}_1u^{(2)}_3-\left(u^{(2)}_2\right)^{2}+u^{(3)}_1u^{(2)}_1u^{(2)}_2+u^{(3)}_2u^{(2)}_2\right.\allowdisplaybreaks\notag\\
&\left.\phantom{\frac{1}{}}+u^{(3)}_1u^{(2)}_3-u^{(2)}_4+u^{(3)}_4\right),\allowdisplaybreaks\\
&\eta_{3}=\frac{1}{4}\left(-\left(u^{(2)}_1\right)^{2}u^{(2)}_3-u^{(2)}_1u^{(2)}_4-u^{(2)}_2u^{(2)}_3+u^{(3)}_1u^{(2)}_1u^{(2)}_3+u^{(3)}_2u^{(2)}_3\right.\allowdisplaybreaks\notag\\
&\left.\phantom{\frac{1}{}}+u^{(3)}_1u^{(2)}_4+u^{(3)}_5\right),\allowdisplaybreaks\notag\\
&\eta_{4}=\frac{1}{4}\left(-\left(u^{(2)}_1\right)^{2}u^{(2)}_4-u^{(2)}_2u^{(2)}_4+u^{(3)}_1u^{(2)}_1u^{(2)}_4+u^{(3)}_2u^{(2)}_4+u^{(3)}_6\right),\notag
\end{align}
and the functions $u^{(3)}_\alpha(t,x)$ $(\alpha=1,\ldots,6)$ satisfy the system
\begin{equation}
\label{sys-6}
\boldsymbol{\Delta}_6\equiv
\left\{
\begin{aligned}
&u^{(3)}_{1,t} + u^{(3)}_1u^{(3)}_{1,x} - u^{(3)}_{1,xx} + u^{(3)}_{2,x} = 0 ,\\
&u^{(3)}_{2,t} + u^{(3)}_2u^{(3)}_{1,x} - u^{(3)}_{2,xx} + u^{(3)}_{3,x} = 0 ,\\
&u^{(3)}_{3,t} + u^{(3)}_3u^{(3)}_{1,x} - u^{(3)}_{3,xx} + u^{(3)}_{4,x} = 0 ,\\
&u^{(3)}_{4,t} + u^{(3)}_4u^{(3)}_{1,x} - u^{(3)}_{4,xx} + u^{(3)}_{5,x} = 0 ,\\
&u^{(3)}_{5,t} + u^{(3)}_5u^{(3)}_{1,x} - u^{(3)}_{5,xx} + u^{(3)}_{6,x} = 0 ,\\
&u^{(3)}_{6,t} + u^{(3)}_6u^{(3)}_{1,x} - u^{(3)}_{6,xx} = 0.
\end{aligned}
\right.
\end{equation}
\end{prop}
\proof
Straightforward, by direct computation.
\endproof

We can repeat the same procedure for the system \eqref{sys-6} made of six coupled Burgers-like equations, and the results are exhibited in the following Proposition.

\begin{prop}
The system \eqref{sys-6}
admits the vector field $\Xi$ of the $Q$-conditional symmetries, say
\begin{equation}
\label{xi-6}
\Xi = \frac{\partial}{\partial t} + \xi(t,x,u^{(3)}_\beta)\frac{\partial}{\partial x} +\sum_{\alpha=1}^6 \eta_{\alpha}(t,x,u^{(3)}_\beta) \frac{\partial}{\partial u^{(3)}_\alpha},
\end{equation}
where
\begin{align}
&\xi=\frac{1}{2}\left(-u^{(3)}_1+u^{(4)}_1\right),\allowdisplaybreaks\notag\\
&\eta_{1}=\frac{1}{4}\left(-\left(u^{(3)}_1\right)^{3}-2u^{(3)}_1u^{(3)}_2+u^{(4)}_1\left(u^{(3)}_1\right)^{2}+u^{(4)}_2u^{(3)}_1\right.\notag\\
&\left.\phantom{\frac{1}{}}+u^{(4)}_1u^{(3)}_2-u^{(3)}_3+u^{(4)}_3\right),\allowdisplaybreaks\notag\\
& \eta_{2}=\frac{1}{4}\left(-\left(u^{(3)}_1\right)^{2}u^{(3)}_2-u^{(3)}_1u^{(3)}_3-\left(u^{(3)}_2\right)^{2}+u^{(4)}_1u^{(3)}_1u^{(3)}_2+u^{(4)}_2u^{(3)}_2\right.\allowdisplaybreaks\notag\\
&\left.\phantom{\frac{1}{}}+u^{(4)}_1u^{(3)}_3-u^{(3)}_4+u^{(4)}_4\right),\allowdisplaybreaks\notag\\
&\eta_{3}=\frac{1}{4}\left(-\left(u^{(3)}_1\right)^{2}u^{(3)}_3-u^{(3)}_1u^{(3)}_4-u^{(3)}_2u^{(3)}_3+u^{(4)}_1u^{(3)}_1u^{(3)}_3+u^{(4)}_2u^{(3)}_3\right.\allowdisplaybreaks\notag\\
&\left.\phantom{\frac{1}{}}+u^{(4)}_1u^{(3)}_4-u^{(3)}_5+u^{(4)}_5\right),\allowdisplaybreaks\\
&\eta_{4}=\frac{1}{4}\left(-\left(u^{(3)}_1\right)^{2}u^{(3)}_4-u^{(3)}_1u^{(3)}_5-u^{(3)}_2u^{(3)}_4+u^{(4)}_1u^{(3)}_1u^{(3)}_4+u^{(4)}_2u^{(3)}_4\right.\allowdisplaybreaks\notag\\
&\left.\phantom{\frac{1}{}}+u^{(4)}_1u^{(3)}_5-u^{(3)}_6+u^{(4)}_6\right),\allowdisplaybreaks\notag\\
&\eta_{5}=\frac{1}{4}\left(-\left(u_1^{(3)}\right)^{2}u^{(3)}_5-u^{(3)}_1u^{(3)}_6-u^{(3)}_2u^{(3)}_5+u^{(4)}_1u^{(3)}_1u^{(3)}_5+u^{(4)}_2u^{(3)}_5\right.\allowdisplaybreaks\notag\\
&\left.\phantom{\frac{1}{}}+u^{(4)}_1u^{(3)}_6+u^{(4)}_7\right),\allowdisplaybreaks\notag\\
&\eta_{6}=\frac{1}{4}\left(-\left(u^{(3)}_1\right)^{2}u^{(3)}_6-u^{(3)}_2u^{(3)}_6+u^{(4)}_1u^{(3)}_1u^{(3)}_6+u^{(4)}_2u^{(3)}_6+u^{(4)}_8\right),\allowdisplaybreaks\notag
\end{align}
and the functions $u^{(4)}_\alpha(t,x)$ $(\alpha=1,\ldots,8)$ satisfy the system
\begin{equation}
\boldsymbol{\Delta}_8 \equiv 
\left\{
\begin{aligned}
&u^{(4)}_{1,t} + u^{(4)}_1u^{(4)}_{1,x} - u^{(4)}_{1,xx} + u^{(4)}_{2,x} = 0,\\
&u^{(4)}_{2,t} + u^{(4)}_2u^{(4)}_{1,x} - u^{(4)}_{2,xx} + u^{(4)}_{3,x} = 0,\\
&u^{(4)}_{3,t} + u^{(4)}_3u^{(4)}_{1,x} - u^{(4)}_{3,xx} + u^{(4)}_{4,x} = 0,\\
&u^{(4)}_{4,t} + u^{(4)}_4u^{(4)}_{1,x} - u^{(4)}_{4,xx} + u^{(4)}_{5,x} = 0,\\
&u^{(4)}_{5,t} + u^{(4)}_5u^{(4)}_{1,x} - u^{(4)}_{5,xx} + u^{(4)}_{6,x} = 0,\\
&u^{(4)}_{6,t} + u^{(4)}_6u^{(4)}_{1,x} - u^{(4)}_{6,xx} + u^{(4)}_{7,x} = 0,\\
&u^{(4)}_{7,t} + u^{(4)}_7u^{(4)}_{1,x} - u^{(4)}_{7,xx} + u^{(4)}_{8,x}= 0,\\
&u^{(4)}_{8,t} + u^{(4)}_8u^{(4)}_{1,x} - u^{(4)}_{8,xx} = 0.
\end{aligned}
\right.
\end{equation}

\end{prop}
\proof
Straightforward, by direct computation.
\endproof

The results heretofore obtained can be summarized as follows:
\begin{itemize} 
\item there are $Q$-conditional symmetries of the system $\boldsymbol{\Delta}_2$ made of two coupled Burgers-like equations expressed in terms of four functions representing arbitrary solutions of the system $\boldsymbol{\Delta}_4$ made of four coupled Burgers-like equations;
\item there are $Q$-conditional symmetries of $\boldsymbol{\Delta}_4$ expressed in terms of six functions representing arbitrary solutions of the system $\boldsymbol{\Delta}_6$ made of six coupled Burgers-like equations; 
\item there are $Q$-conditional symmetries of $\boldsymbol{\Delta}_6$ expressed in terms of eight functions representing arbitrary solutions of the system $\boldsymbol{\Delta}_8$ made of eight coupled Burgers-like equations.
\end{itemize}

These results suggest to conjecture that repeatedly searching for $Q$-condi\-tional symmetries, and starting from a pair of coupled Burgers-like equations, a hierarchy of systems made of an even number of coupled Burgers-like equations arises.

Indeed, the latter conjecture and the one made in the previous Subsection, can be unified and proved to be true, as it will be shown in the following Section. 

\section{The general hierarchy of Burgers-like\\ equations}
\label{sec:4}
In this Section, we show that the existence of both hierarchies of Burgers-like equations, arising by searching at each step non trivial $Q$-conditional symmetries, can be proved in general. In fact, we have an infinite hierarchy of systems made of an odd number of coupled Burgers-like equations or made of an even number of coupled Burgers-like equations depending on the starting point. The results obtained in 
Subsections~\ref{subsec:3.1} and \ref{subsec:3.2} for the first members of the two hierarchies are useful for guessing the structure of the vector fields of the $Q$-conditional symmetries for the generic element of the hierarchy. This allows to state the following Theorem.
 
\begin{thm}
\label{thm:main}
Let $m$ be a positive integer, and let $k = \lceil m/2 \rceil$.  The system of Burgers-like equations
\begin{equation}
\label{eqn:BurgersLikeN}
\boldsymbol\Delta_m\equiv\left\{
\begin{aligned}
&u^{(k)}_{\alpha,t}+u^{(k)}_\alpha u^{(k)}_{1,x}-u^{(k)}_{\alpha,xx}+u^{(k)}_{\alpha+1,x}=0,\\
&u^{(k)}_{m,t}+u^{(k)}_m u^{(k)}_{1,x}-u^{(k)}_{m,xx}=0,
\end{aligned}
\right.
\end{equation}
with $\alpha=1,\ldots,m-1$, admits the $Q$-conditional symmetries associated to the vector field
\begin{equation}
\label{xi-n}
\Xi = \frac{\partial}{\partial t} + \xi(t,x,u^{(k)}_\beta)\frac{\partial}{\partial x} +\sum_{\alpha=1}^m \eta_{\alpha}(t,x,u^{(k)}_\beta) \frac{\partial}{\partial u^{(k)}_\alpha},
\end{equation}
where
\begin{equation}
\label{eqn:infgensN}
\begin{aligned}
&\xi=\frac{1}{2}\left(-u^{(k)}_{1}+u^{(k+1)}_1\right),\\
&\eta_{\alpha}=\frac{1}{4}\left(-\left(u^{(k)}_1\right)^2 u^{(k)}_\alpha - u^{(k)}_1 u^{(k)}_{\alpha+1} - u^{(k)}_2 u^{(k)}_\alpha + u^{(k+1)}_1 u^{(k)}_1 u^{(k)}_\alpha  \right.\\
&\left.\phantom{\frac{1}{}}+ u^{(k+1)}_2 u^{(k)}_\alpha+u^{(k+1)}_1 u^{(k)}_{\alpha+1} - u^{(k)}_{\alpha+2} + u^{(k+1)}_{\alpha+2}\right),\\
&\eta_{m-1}=\frac{1}{4}\left(-\left(u^{(k)}_1\right)^2 u^{(k)}_{m-1} - u^{(k)}_1 u^{(k)}_m - u^{(k)}_2 u^{(k)}_{m-1} + u^{(k+1)}_1 u^{(k)}_1 u^{(k)}_{m-1}\right.\\
&\left.\phantom{\frac{1}{}}+ u^{(k+1)}_2 u^{(k)}_{m-1} + u^{(k+1)}_1 u^{(k)}_m + u^{(k+1)}_{m+1}\right),\\
&\eta_{m}=\frac{1}{4}\left(-\left(u^{(k)}_1\right)^2 u^{(k)}_m - (1-\delta_{1m})u^{(k)}_2 u^{(k)}_m + u^{(k+1)}_1 u^{(k)}_1 u^{(k)}_m \right.\\
&\left.\phantom{\frac{1}{}}+ u^{(k+1)}_2 u^{(k)}_m + u^{(k+1)}_{m+2}\right),
\end{aligned}
\end{equation}
$\delta_{1m}$ being the Kronecker symbol,
with $\alpha=1,\dots,m-2$, provided that the functions $u^{(k+1)}_\alpha(t,x)$ satisfy the system
\begin{equation}
\label{eqn:compatibilityN}
\boldsymbol{\Delta}_{m+2}\equiv
\left\{
\begin{aligned}
&u^{(k+1)}_{\alpha,t}+u^{(k+1)}_\alpha u^{(k+1)}_{1,x}-u^{(k+1)}_{\alpha,xx}+u^{(k+1)}_{\alpha+1,x}=0,\\
&u^{(k+1)}_{m+2,t}+u^{(k+1)}_{m+2}u^{(k+1)}_{1,x}-u^{(k+1)}_{m+2,xx}=0,
\end{aligned}
\right.
\end{equation}
with $\alpha=1,\ldots,m+1$. 
\end{thm}
\proof
It must be verified that the vector field \eqref{xi-n}, defined by  \eqref{eqn:infgensN}, is admitted by the system (\ref{eqn:BurgersLikeN}) along with the constraints \eqref{eqn:compatibilityN}. 
In fact, requiring
\begin{equation}
\left.\Xi^{(2)}\left(\boldsymbol{\Delta}_m\right)\right|_{\mathcal{M}_m}=\mathbf{0},
\end{equation}
where the manifold $\mathcal{M}_m$ of the jet space is defined by 
\begin{equation}
\left\{
\begin{aligned}
&\boldsymbol{\Delta}_m=\mathbf{0},\\
&Q_\alpha\equiv \Xi\left(u^{(k)}_\alpha- u^{(k)}_{\alpha}(t,x)\right)=0,\qquad \alpha=1,\ldots,m,\\
&\frac{DQ_\alpha}{Dt} = \frac{DQ_\alpha}{Dx}=0,
\end{aligned}
\right.
\end{equation} 
we get the following polynomial system of $m$ equations in the variables $u^{(k)}_\alpha$ and  $u^{(k)}_{\alpha,x}$:
\begin{equation}
\label{inv:polynomial}
\begin{aligned}
& \left(u_{1,t}^{(k+1)} + u_1^{(k+1)} u_{1,x}^{(k+1)} - u_{1,xx}^{(k+1)} + u_{2,x}^{(k+1)}\right) u_1^{(k)} u_\alpha^{(k)} \\
&\quad+ \left(u_{2,t}^{(k+1)} + u_2^{(k+1)} u_{1,x}^{(k+1)} - u_{2,xx}^{(k+1)} + u_{3,x}^{(k+1)}\right) u_\alpha^{(k)} \\
&\quad+\left(u_{1,t}^{(k+1)} + u_1^{(k+1)} u_{1,x}^{(k+1)} - u_{1,xx}^{(k+1)} + u_{2,x}^{(k+1)}\right) u_{\alpha+1}^{(k)}\\
&\quad-2\left(u_{1,t}^{(k+1)} + u_1^{(k+1)} u_{1,x}^{(k+1)} - u_{1,xx}^{(k+1)} + u_{2,x}^{(k+1)}\right) u_{\alpha,x}^{(k)} \\
&\quad+ u_{\alpha+2,t}^{(k+1)} + u_{\alpha+2}^{(k+1)} u_{1,x}^{(k+1)} - u_{\alpha+2,xx}^{(k+1)} + u_{\alpha+3,x}^{(k+1)} = 0,\\
& \left(u_{1,t}^{(k+1)} + u_1^{(k+1)} u_{1,x}^{(k+1)} - u_{1,xx}^{(k+1)} + u_{2,x}^{(k+1)}\right) u_1^{(k)} u_m^{(k)} \\
&\quad+ \left(u_{2,t}^{(k+1)} + u_2^{(k+1)} u_{1,x}^{(k+1)} - u_{2,xx}^{(k+1)} + u_{3,x}^{(k+1)}\right) u_m^{(k)}\\
&\quad-2\left(u_{1,t}^{(k+1)} + u_1^{(k+1)} u_{1,x}^{(k+1)} - u_{1,xx}^{(k+1)} + u_{2,x}^{(k+1)}\right) u_{m,x}^{(k)}\\
&\quad+ u_{m+2,t}^{(k+1)} + u_{m+2}^{(k+1)} u_{1,x}^{(k+1)} - u_{m+2,xx}^{(k+1)} = 0,
\end{aligned}
\end{equation}
where $\alpha=1,\dots,m-1$.

Due to the arbitrariness of  $u^{(k)}_\alpha$ and  $u^{(k)}_{\alpha,x}$, the system \eqref{inv:polynomial} is satisfied if and only if 
\begin{equation}
\left\{
\begin{aligned}
&u_{1,t}^{(k+1)} + u_1^{(k+1)} u_{1,x}^{(k+1)} - u_{1,xx}^{(k+1)} + u_{2,x}^{(k+1)}=0,\\
&u_{2,t}^{(k+1)} + u_2^{(k+1)} u_{1,x}^{(k+1)} - u_{2,xx}^{(k+1)} + u_{3,x}^{(k+1)}=0,\\
&u_{\alpha+2,t}^{(k+1)} + u_{\alpha+2}^{(k+1)} u_{1,x}^{(k+1)} - u_{\alpha+2,xx}^{(k+1)} + u_{\alpha+3,x}^{(k+1)}=0,\quad \alpha=1,\dots,m-1,\\
&u_{m+2,t}^{(k+1)} + u_{m+2}^{(k+1)} u_{1,x}^{(k+1)} - u_{m+2,xx}^{(k+1)}=0,
\end{aligned}
\right.
\end{equation}
\emph{i.e.}, the system $\boldsymbol{\Delta}_{m+2}=\mathbf{0}$ has to be satisfied.
\endproof

\begin{remark}
Note that if $m$ is odd (even, respectively), a hierarchy of systems with an odd (even, respectively) number of equations is generated.
\end{remark}

Each element of the infinite hierarchy of systems of Burgers-like equations can be written in the form of a matrix Burgers' equation \cite{Mansfield1999,Arrigo-Hickling} that can be linearized by means of the matrix Hopf-Cole transformation \cite{Levi1983}.

In fact, defining the $m\times m$ matrix $\Omega$  as
\begin{equation}
\Omega = \left[
\begin{array}{ccccc}
0 & 1 & \cdots & 0 & 0\\
0 & 0 & 1 & \cdots & 0\\
\vdots & \vdots & \vdots & \vdots & \vdots \\
0 & 0 & 0 & \cdots & 1\\
u^{(k)}_{m} & u^{(k)}_{m-1} & \cdots & u^{(k)}_{2} & u^{(k)}_{1} 
\end{array}
\right],
\end{equation}
the system \eqref{eqn:BurgersLikeN} writes in the form of a matrix Burgers' equation, say
\begin{equation}
\label{matrixBurgers}
\Omega_{,t}+\Omega_{,x}\Omega-\Omega_{,xx}=0.
\end{equation}
The matrix Hopf-Cole transformation \cite{Levi1983,Arrigo-Hickling},
\begin{equation}
\label{matrixHopfCole}
\Omega = -2\Phi_{,x}\Phi^{-1},
\end{equation}
$\Phi$ being an invertible $m\times m$ matrix with entries depending on $t$ and $x$,
maps \eqref{matrixBurgers} to a matrix heat equation,
\begin{equation}
\Phi_{,t}-\Phi_{,xx}=0;
\end{equation}
moreover, from \eqref{matrixHopfCole}, it results $\Phi_{,x}=-\frac{1}{2}\Omega\Phi$, whereupon, computing the entries of $\Omega\Phi$,  the solution of the system \eqref{eqn:BurgersLikeN} is achieved from the linear algebraic system
\[
\left\{
\begin{aligned}
&u^{(k)}_{m}v_1+\sum_{j=1}^{m-1}(-2)^j u^{(k)}_{m-j}
\frac{\partial^j v_1}{\partial x^j}=(-2)^{m}\frac{\partial^m v_1}{\partial x^m},\\
&u^{(k)}_{m}v_2+\sum_{j=1}^{m-1}(-2)^j u^{(k)}_{m-j}
\frac{\partial^j v_2}{\partial x^j}=(-2)^{m}\frac{\partial^m v_2}{\partial x^m},\\
&\cdots,\\
&u^{(k)}_{m}v_m+\sum_{j=1}^{m-1}(-2)^j u^{(k)}_{m-j}
\frac{\partial^j v_m}{\partial x^j}=(-2)^{m}\frac{\partial^m v_m}{\partial x^m},
\end{aligned}
\right.
\]
where $v_\alpha(t,x)$ $(\alpha=1,\ldots,m)$ are $m$ solutions of linear heat equations, \emph{i.e.}, 
\[
v_{\alpha,t}-v_{\alpha,xx}=0, \qquad \alpha=1,\ldots,m.
\]
Therefore, similarly to what happens for classical Burgers' equation, also the solutions of each element of the hierarchy of systems of coupled Burgers-like equations, because of the matrix Hopf-Cole transformation, can be obtained by the solutions of linear heat equation. 

As a last comment, we observe that the classical Lie point symmetries of a generic element of this infinite hierarchy span a five-dimensional Lie algebra, as shown below.

\begin{prop}
Let $m$ be a positive integer, and let $k = \lceil m/2 \rceil$.  The system of Burgers-like equations
\begin{equation}
\label{eqn:BurgersLikeN-bis}
\boldsymbol\Delta_m\equiv
\left\{
\begin{aligned}
&u^{(k)}_{\alpha,t}+u^{(k)}_\alpha u^{(k)}_{1,x}-u^{(k)}_{\alpha,xx}+u^{(k)}_{\alpha+1,x}=0,\qquad \alpha=1,\ldots,m-1,\\
&u^{(k)}_{m,t}+u^{(k)}_m u^{(k)}_{1,x}-u^{(k)}_{m,xx}=0,
\end{aligned}
\right.
\end{equation} 
for $m=1$ (classical Burgers' equation) admits the Lie point symmetries generated by:
\begin{equation*}
\begin{aligned}
&\Xi_1=\frac{\partial}{\partial t}, \qquad \Xi_2 = \frac{\partial}{\partial x},\\
&\Xi_3=2t\frac{\partial}{\partial t}+x\frac{\partial}{\partial x}-u^{(1)}_1\frac{\partial}{\partial u^{(1)}_1},\\
&\Xi_4=t\frac{\partial}{\partial x}+\frac{\partial}{\partial u^{(1)}_1},\\
&\Xi_5 = t^2\frac{\partial}{\partial t}+tx\frac{\partial}{\partial x}+(x-tu^{(1)}_1)\frac{\partial}{\partial u^{(1)}_1};
\end{aligned}
\end{equation*}
for $m=2$ the Lie point symmetries generated by:
\begin{equation*}
\begin{aligned}
&\Xi_1=\frac{\partial}{\partial t}, \qquad \Xi_2 = \frac{\partial}{\partial x},\\
&\Xi_3=2t\frac{\partial}{\partial t}+x\frac{\partial}{\partial x}-u^{(1)}_1\frac{\partial}{\partial u^{(1)}_1}-2u^{(1)}_2\frac{\partial}{\partial u^{(1)}_2},\\
&\Xi_4=t\frac{\partial}{\partial x}+2\frac{\partial}{\partial u^{(1)}_1}
-u^{(1)}_1\frac{\partial}{\partial u^{(1)}_2},\\
&\Xi_5 = t^2\frac{\partial}{\partial t}+tx\frac{\partial}{\partial x}+(2x-tu^{(1)}_1)\frac{\partial}{\partial u^{(1)}_1} -(xu^{(1)}_1+2tu^{(1)}_2+2)\frac{\partial}{\partial u^{(1)}_2};
\end{aligned}
\end{equation*}
for $m\ge 3$ the Lie point symmetries generated by:
\begin{equation*}
\begin{aligned}
\Xi_1&=\frac{\partial}{\partial t}, \qquad \Xi_2 = \frac{\partial}{\partial x},\\
\Xi_3&=2t\frac{\partial}{\partial t}+x\frac{\partial}{\partial x}-\sum_{\alpha=1}^m\alpha u^{(k)}_\alpha\frac{\partial}{\partial u^{(k)}_\alpha},\\
\Xi_4&=t\frac{\partial}{\partial x}+m\frac{\partial}{\partial u^{(k)}_1}
+\sum_{\alpha=2}^m (\alpha-m-1)u^{(k)}_{\alpha-1}\frac{\partial}{\partial u^{(k)}_\alpha},\\
\Xi_5 &= t^2\frac{\partial}{\partial t}+tx\frac{\partial}{\partial x}+\left(mx-tu^{(k)}_1\right)\frac{\partial}{\partial u^{(k)}_1}\\
&-\left((m-1)(xu^{(k)}_1+m) +2tu^{(k)}_2\right)\frac{\partial}{\partial u^{(k)}_2} \\
&- \sum_{\alpha=3}^m \left(\alpha t u^{(k)}_\alpha+(m-\alpha+1)\left(xu^{(k)}_{\alpha-1}-(m-\alpha+2)u^{(k)}_{\alpha-2}\right)\right)\frac{\partial}{\partial u^{(k)}_\alpha}.
\end{aligned}
\end{equation*}
Whatever the number $m$ of coupled equations is, we have always a five-dimensional Lie algebra (time and space translation, scaling, Galilean and projective transformation, respectively); these Lie algebras,  although realized in terms of vector fields on manifolds with different dimensionality, share the same structure constants and so they are all isomorphic.
\end{prop}

\section{Conclusions}
\label{sec:5}
In this paper, repeatedly searching for $Q$-conditional symmetries, we derived an infinite hierarchy of coupled Burgers-like equations. After fixing the notation and shortly reviewing the approach to $Q$-conditional symmetries, we analyzed some classes of Burgers-like equations and determined their nonclassical symmetries. At first, the $Q$-conditional symmetries of Burgers-like systems made by one, three and five differential equations have been considered. The admitted $Q$-conditional symmetries suggest the existence of a chain of systems made of an odd number of Burgers-like equations. The same behavior occurs by considering an even number of Burgers-like equations. Then, we proved a theorem that merges both hierarchies, providing the existence of nonclassical symmetries of each element in the hierarchy. 
Moreover, the elements of this hierarchy can be written as a matrix Burgers' equation that is linearized by the matrix Hopf-Cole transformation; as a consequence, the general solution can be obtained in terms of solutions of linear heat equations. As a last result, it is shown that the Lie algebra of point symmetries admitted by each element of this infinite hierarchy is five-dimensional and isomorphic to the Lie algebra admitted by classical Burgers' equation.

\section*{Acknowledgments}
This work was supported by PRIN 2022 PNRR P20222B5P9 ``Non linear models for magma 
transport and volcanoes generation'', and by ``Gruppo Nazionale per la Fisica Matematica'' (GNFM) of 
``Istituto Nazionale di Alta Matematica''.


\begin{thebibliography}{99}

\bibitem{Lie1} Lie, S., Engel, F.: Theorie der transformationsgruppen. Teubner: Leipzig, Germany (1888)

\bibitem{Lie2} Lie, S.: Vorlesungen \"{u}ber differentialgleichungen mit bekannten infinitesimalen Transformationen. Teubner: Leipzig, Germany (1891)

\bibitem{Ovsiannikov} Ovsiannikov, L.V.: Group analysis of differential equations. Academic Press, New York  (1982)

\bibitem{Olver1} Olver, P.J.: Applications of Lie groups to differential equations. Springer, New York (1986)

\bibitem{Bluman-Kumei} G.~W.~Bluman, S.~Kumei, Symmetries and differential equations, Springer, New York, 1989.

\bibitem{Ibragimov:CRC} Ibragimov, N.H.: Editor, CRC Handbook of Lie group analysis of differential equations (three volumes). CRC Press, Boca Raton (1994, 1995, 1996)

\bibitem{Hereman1997}
Hereman, W.: Review of symbolic software for Lie symmetry analysis.
Math. Comput. Model. {\bf  25}, 115--132 (1997)

\bibitem{Baumann2000book} Baumann, G.: Symmetry analysis of differential equations with 
Mathematica. Springer, New York (2000)

\bibitem{CarminatiWu2000} Carminati, J., Vu, K.~T.:
Symbolic computation and differential equations: Lie symmetries.
J. Symb. Comput. {\bf  29}, 95--116 (2000) 


\bibitem{ButcherCarminati2003}
Butcher, J., Carminati, J., Vu, K.~T.:
A comparative study of some computer algebra packages which determine
the Lie point symmetries of differential equations.
Comput. Phys. Commun.  {\bf 155}, 92--114 (2003)

\bibitem{Cheviakov2007}
Cheviakov, A.~F.: GeM software package for computation of symmetries and 
conservation laws of differential equations. Comput. Phys. Commun.  {\bf 176}, 48--61 (2007)

\bibitem{Steeb2007book}
Steeb, W.~H.: Continuous Symmetries, Lie algebras, Differential Equations
and Computer Algebra (2nd ed.). World Scientific Publishing, Singapore (2007)

\bibitem{Cheviakov2010}
Cheviakov, A.~F.: Symbolic computation of local symmetries of nonlinear and linear
  partial and ordinary differential equations.
 Math. Comput. Sci. {\bf  4}, 203--222 (2010)

\bibitem{Sade2010}
Rocha~Filho, T.~M., Figueiredo, A.:
 [SADE] a Maple package for the symmetry analysis of differential equations.
 {\em Comput. Phys. Commun.}  {\bf 182}, 467--476 (2010)

\bibitem{JeffersonCarminati2013}
Jefferson, G.~F., Carminati, J.: ASP: Automated symbolic computation of approximate symmetries of  differential equations. Comput. Phys. Commun.  {\bf 184}, 1045--1063 (2013)

\bibitem{Oliveri:relie} Oliveri, F.: ReLie: a Reduce program for Lie group analysis of differential equations. Symmetry \textbf{13}, 1--39 (2021) 

\bibitem{Reduce} Hearn, A.~C., Sch\"opf, R.: Reduce Users' Manual. Free Version, available online: https://reduce-algebra.sourceforge.io (2024).

\bibitem{BlumanAnco} Bluman, G.~W., Anco, S.~C.: Symmetry and integration methods for differential equations. Springer, New York (2002)

\bibitem{Olver-Rosenau-2} Olver, P.~J.,  Rosenau, P.: Group invariant solutions of differential equations, SIAM J. Appl. Math. \textbf{47}, 263--278  (1987)

\bibitem{Noether} E.~Noether, Invariante Variationsprobleme. Kgl. Ges. d. Wiss. Nachrichten, Math.-phys. Klasse,  235--257 (1918).

\bibitem{BlumanCheviakovAnco} Bluman, G.~W.,  Cheviakov, A.~F.,  Anco, S.~C.: Applications of symmetry methods to partial differential equations. Springer, New York (2009)

\bibitem{Oliveri:Symmetry} Oliveri, F.: Lie symmetries of differential equations: classical results and recent contributions. Symmetry \textbf{2}, 658--706 (2010) 

\bibitem{Oliveri:quasilinear} Oliveri, F.: General dynamical systems described by first order quasilinear PDEs reducible to homogeneous and autonomous form. Int. J. Non--Linear Mech. \textbf{47},  53--60 (2012)

\bibitem{GorgoneOliveri2}
Gorgone, M., Oliveri, F.: Nonlinear first order PDEs reducible to autonomous form polynomially homogeneous in the derivatives. J. Geom. Phys. \textbf{113}, 53--64 (2017) 

\bibitem{Bluman-Cole-1969} Bluman, G.~W., Cole, J.~D.: The general similarity solution of the heat equation. J. Math. Mech. \textbf{18}, 1025--1042 (1969) 

\bibitem{Fuschich} Fushchich, W.~I.: How to extend symmetry of differential equation?, in: Symmetry and Solutions of Nonlinear Equations of Mathematical Physics, Inst. Math. Acad. Sci. Ukra.,  4--16 (1987)

\bibitem{Fuschich-Tsyfra} Fushchych, W.~I., Tsyfra, I.~M.: On a reduction and solutions of the nonlinear wave equations with broken symmetry. J. Phys. A: Math. Gen. \textbf{20}, L45--L48 (1987) 

\bibitem{Levi-Winternitz} Levi, D., Winternitz, P.: Nonclassical symmetry reduction: example of the Boussinesq equation. J. Phys. A: Math. Gen. \textbf{22}, 2915--2924 (1989) 

\bibitem{Clarkson-Kruskal} Clarkson, P.~A., Kruskal, M.: New similarity reductions of the Boussinesq equation. J. Math. Phys. \textbf{30},  2201--2213 (1989)

\bibitem{PucciSaccomandi} Pucci, E., Saccomandi, G.: On the weak symmetry groups of partial fifferential equations. J. Math. Anal. Appl. \textbf{163},  588--598 (1992)

\bibitem{Arrigo} Arrigo, D.~J., Broadbridge, P., Hill, J.~M.: Nonclassical symmetry solutions and the methods of Bluman--Cole and Clarkson--Kruskal. J. Math. Phys. \textbf{34}, 4692--4703 (1993)

\bibitem{Olver1994} Olver, P.~J.: Direct reduction and differential constraints. Proc. Roy. Soc. London Ser. A \textbf{444}, 509--523 (1994) 
.
\bibitem{Olver-Vorobev}
Olver, P.~J., Vorob'ev, E.~M.: Nonclassical and conditional symmetries. In: CRC Handbook of Lie group analysis of differential equations  \textbf{3}, 291--328  (1996)

\bibitem{Saccomandi} Saccomandi, G.: A personal overview on the reduction methods for partial differential equations, Note Mat. \textbf{23}, 217--248 (2004/2005)	

\bibitem{Cherniha-JMAA2007} Cherniha, R.: New Q-conditional symmetries and exact solutions of some reaction--diffusion--convection equations arising in mathematical biology. J. Math. Anal. Appl. \textbf{326}, 783--799 (2007) 

\bibitem{Cherniha-JPA2010} Cherniha, R.: Conditional symmetries for systems of PDEs: new definitions and their application for reaction--diffusion systems. J. Phys. A: Math. Theor. \textbf{41}, 405207 (2010) 

 \bibitem{Cherniha-Davydovych} Cherniha, R., Davydovych, V.: Nonlinear reaction-diffusion systems. Conditional symmetries, Exact solutions and their applications in biology, Lecture Notes in Mathematics, \textbf{2196}, Springer, Cham  (2017)

\bibitem{Kunzinger-Popovich} Kunzinger M., Popovych R.~O.: Is a nonclassical symmetry a symmetry? Proceedings of 4th Workshop ``Group Analysis of Differential Equations and Integrable Systems'', University of Cyprus, Nicosia, 107--120 (2009) 

\bibitem{Crighton} Crighton, D.~G.: Basic theoretical nonlinear acoustics, Frontiers in
Physical Acoustics, XCIII course Italian Physical Society, Bologna (1986)

\bibitem{Hopf} Hopf, E.: The partial differential equation $u_{t}+uu_{x}-\mu u_{xx}= 0$. Comm. Pure Appl. Math. \textbf{3}, 201--230 (1950) 

\bibitem{Cole} Cole, J.~D.: On a quasilinear parabolic equation occurring
in aerodynamics. Quart. Appl. Math. \textbf{9}, 225--236 (1951) 

\bibitem{Mansfield1999} Mansfield, E.~L.: Nonclassical group analysis of the heat equation. J. Math. Anal. Appl., \textbf{231}, 526--542 (1999) 

\bibitem{Arrigo-Hickling} Arrigo, D.~J., Hickling, F.: On the determining equations for the
nonclassical reductions of the heat and Burgers' equation. 
J. Math. Anal. Appl. \textbf{270}, 582--589 (2002) 

\bibitem{Arrigo-Ekrut} Arrigo, D.~J., Ekrut, D.~A., Fliss, J.~R., Le, L.: 
Nonclassical symmetries of a class of Burgers' systems. J. Math. Anal. Appl. \textbf{371},  813--820 (2010)

\bibitem{Cherniha-Serov}
Cherniha, R., Serov, M.: Nonlinear systems of the Burgers-type equations: Lie and Q-conditional symmetries, Ans\"{a}tze and solutions. J. Math. Anal. Appl. \textbf{282}, 305--328 (2003) 

\bibitem{Webb} Webb, G.~M.: Lie symmetries of a coupled nonlinear Burgers--heat
equation system. J. Phys. A \textbf{23}, 3885--3894 (1990) 

\bibitem{Levi1983} Levi, D., Ragnisco, O., Bruschi, M.: Continuous and discrete 
Burgers’ hierarchies. Nuovo Cimento B \textbf{74}, 33--51 (1983) 

\bibitem{Nucci1994} Nucci, M.~C.: Iterating the nonclassical symmetries method. Physica D: Nonlinear Phenomena \textbf{78}, 124--134 (1994)

\bibitem{Allassia-Nucci} Allassia, F., Nucci, M.~C.: Symmetries and heir equations for the laminar boundary layer model. J. Math. Anal. Appl. \textbf{201}, 911--942 (1996)

\bibitem{Nucci-25years} Nucci, M.~C.: Heir-equations for partial differential equations: a 25-year review. In ``Nonlinear systems and their remarkable mathematical structures''. Vol. 2, 188--205, CRC Press, Boca Raton, (2020)

\end{thebibliography}
\end{document}